\documentstyle[12pt]{article}
\begin{document}

\title {Compatibility of Distortion Fields Caused by
Topological Defects in 2D Lattices}
\author {Dominik Rogula\thanks{Center for Physics and Applied
Mathematics, Institute of Fundamental
Technological Research, ul. Swietokrzyska 21, 00-049 Warsaw, Poland.
E--mail: drogula@ippt.gov.pl}}
\date{November 24, 1998}
\maketitle

\newcommand{\equn}[1]{\begin{equation} {#1} \end{equation}}
\newcommand{\equln}[2]{\begin{equation} {#1} \label{#2} \end{equation}}


 \noindent {\bf Abstract.} -- Topological defects in crystalline
 lattices are considered. In relation to physical realizability
 of such defects, criteria for geometric compatibility of the
 lattice distortions are formulated. For 2D lattices it is shown
 that the answer to the question of existence of distortion fields
 which are both geometrically compatible and homotopically
 non--trivial is in the affirmative.
\normalsize

\section{Introduction}\indent
     Since the paper \cite{Finkelstein 66} which first introduced
homotopy into physics, considerable interest in topologically
non--trivial configurations of physical systems has arisen.
In particular, it concerns the topological classification of
structural defects in ordered materials. A discussion of the subject
may be found
in \cite{Ovidko+Romanov}. Fundamental principles of the
topological classification scheme in the framework of broken
symmetry have been given in \cite{Michel 77}.

     While substantial progress for translationally invariant
continuous media has been achieved (see \cite{Trebin, Michel 80}),
crystalline structures turned out to be more difficult.
Two different approaches have been proposed, based on affine
\cite{rogula 75} and isometric \cite{Michel 77} groups,
respectively. As noted in \cite{Mermin}, in the case of crystalline
medium not every continuous map need correspond to a physical
state. Physically admissible states should be restricted by
appropriate compatibility conditions, which would express the relations
between local states at infinitesimally close points of the medium.
Such conditions might, in principle, render non--trivial homotopy
classes physically not realizable, what would make homotopic
classification ineffective.

     In the present paper we propose appropriate compatibility
conditions for 2D lattices. Although the problem makes
sense also for higher dimensionalities, the 2D case is distinguished,
and deserves separate consideration for several reasons. Firstly,
the space $R^2$ admits natural complex structure, which converts it
into $C^1$, endowed with powerful analytic properties.
Secondly, as is well known from anyonic physics \cite{Wilczek},
the topological covering properties of planar
systems are exceptional and more complicated (braid groups) than for
other space dimensionalities.
Moreover, 2D physical structures are of direct interest in
surface physics, crystalline interfaces, high T$_c$ superconducting
CuO$_2$ planes, and in other physical situations where planar
sub-structures play dominant role. The 2D case provides also a good
point of departure for developing the 3D theory. For reasons of
simplicity, we restrict our present considerations to Bravais lattices.
Topological description of crystalline lattices of complex structure
is given in \cite{Rogula 96}.

\section{The manifold of planar Bravais lattices}\indent

     To discuss the topology of lattice distortions in a reasonably
precise way, one needs a topological space whose points can be put
into a bijective correspondence with the set of all ideal lattice
states \cite{rogula 75}. For 2D lattices it is a 6D topological
manifold $\Lambda$
\equln{ \Lambda \stackrel{\cdot}{=} G/\Gamma}{LaGGa}
where $G = {\rm GA}^+(2,R)$ and $\Gamma = {\rm SA}(2,Z)$.
Taking into account the topology of $G$ ($\Gamma$ is then a
discrete subgroup of $G$) one endows $\Lambda$ with the
quotient topology. In the above construction,
two lattices are considered identical iff they
coincide as subsets of the plane $E^2$.

     We also make use of the standard representation of a lattice
by an affine frame, consisting of a point $a\ \epsilon\ E^2$
and two base vectors $a_K$, $K=1,2$. It is, however, crucial
to observe that the space of affine frames (which will be
here identified with the affine group  GA$(2,R)$)
is not equivalent to $\Lambda$. This
is due to the well known fact that, while an affine frame
determines a unique lattice, the converse does not hold.
Two different frames can determine the same lattice.

   This defines an equivalence relation between the affine frames.
Let
\equln{r^{(1)}=(a^{(1)},a^{(1)}_{\ 1},a^{(1)}_{\ 2}),
   \ \ r^{(2)}=(a^{(2)},a^{(2)}_{\ 1},a^{(2)}_{\ 2})}{framesr}
be  a  pair of such frames with the same (positive) orientation.
Then $r^{(1)}$ and $r^{(2)}$ are  equivalent iff there exists
a unimodular matrix $A_K^{\ L}$ and a column $b^L$,
both of integer components and such that
\begin{eqnarray}
  a^{(2)}_{\ K}& = &A_K^{\ L} a^{(1)}_{\ L},\nonumber\\
  b^{(2)}&=&b^{(1)}+b^L a^{(1)}_{\ L}.
\end{eqnarray}
In shorthand it can be put as
\equn{r^{(2)}= \gamma r^{(1)},\ \ \ \gamma\ \epsilon\ \Gamma.}

     By a reference lattice we shall understand the lattice which
under (\ref{LaGGa}) corresponds to the unit matrix in $G$.
The actual form of this lattice depends on the reference frame.

     Let us briefly discuss the global structure of the manifold
$\Lambda$. The notation SA$(2,R)$ will be shortened here to $H$.

     1. $\Lambda$ can be represented as the topological product
\equn{\Lambda = R_+\times\Sigma,\ \ \Sigma = H/ \Gamma.}
The manifold $\Sigma$ is composed of the lattices that
are equi-areal with the reference lattice.
$\Lambda$ and $\Sigma$ have identical homotopy type.

     2. $\Lambda$ (resp. $\Sigma$) is topologically covered by
$G$ (resp. $H$). In both cases the
fibre of the covering coincides with $\Gamma$.
In consequence $\Lambda$ (or $\Sigma$) and $G$ (or $H$)
have identical homotopy properties for dimensions $n\geq 2$.
The coverings carry the structure of differentiable manifold from
the Lie groups $G$, $H$ to the spaces $\Lambda$, $\Sigma$.

     3. $\Lambda$ is a homogeneous $G$--space, $\Sigma$
is a homogeneous $H$--space. In both cases
the stationary subgroup of the reference lattice equals $\Gamma$.

     4. The action of the Euclidean isometries $E\subset G$
on $\Lambda$ determines a fibration
\equn{p: \Lambda\rightarrow \Lambda/E}
where $\Lambda/{\rm E}$ denotes the relevant space of orbits.
The fibration $p$ does not, however, satisfy the bundle property.
Similiar construction holds for $E$, $H$ and $\Sigma$.

     The structure of the space of orbits $\Lambda/E$
can be investigated by making
use of the results given by Gruber \cite{Gruber}.
The four strata of $\Lambda/E$ correspond
to the four 2D crystallographic systems.

\section{Compatibility of the distortion fields}\indent

Let $B$ be a 2D region occupied by a planar material structure and let
\equln{f: B\rightarrow \Lambda}{eqfBLa}
be the distortion field in $B$. For our present purposes it is
sufficient to consider connected open regions $B$ of the form
 \equln{B = {\rm int}\ A\ \setminus\ \bigcup_{\iota\,\epsilon\,J}
            A_\iota}{eqsetB}
where $\iota$ runs through a finite index set $J$, while $A$
and $A_\iota$ are contractible compact subsets of $E^2$,
$A_\iota$'s being mutually disjoint subset of ${\rm int}\ A$.
The covering $G\rightarrow \Lambda$
guarantees then the existence of an open covering
$\{B_\alpha\}$ of $B$ and a collection of local fields
\equn{ g_\alpha: B_\alpha\rightarrow G}
which satisfy the equivalence relation
\equln{ g_\beta (x)\ \epsilon\ \Gamma g_\alpha (x)
      \ \ \ {\rm for}\ \ x\ \epsilon\ B_\alpha\cap B_\beta}{eqvgGa}
and correspond to the field (\ref{eqfBLa}). In this way the problem
of geometric compatibility of a distortion field (\ref{eqfBLa})
reduces locally to the problem of
geometric compatibility of appropriate fields of affine frames,
which can be solved by differential-geometric methods.

     To that effect we shall apply locally the approach
presented in \cite{Mistura}. A differentiable field of affine
frames defines a soldering form $\theta$ and a field of linear
coframes $\omega$ which together determine an affine connection
which, projected into linear connections, is a
teleparallelism. The corresponding affine curvature is
composed of the translation part
\equln{D\theta = d\theta + \omega \wedge \theta}{Dtheta}
and the linear part
\equln{D\omega = d\omega + \frac{1}{2}[\,\omega, \omega\,] = 0}{Dw}
(temporarily we suppress the index $\alpha$).
$D$ stands here for the covariant exterior differentiation defined
by our connection.

While the linear curvature $D\omega$ vanishes identically,
the translation part $D\theta$ in general does not vanish.
In geometric terms it defines the torsion $\tau = D\omega$
which, in turn, may be
interpreted physically as the density of a certain
continuous distribution of dislocations.
In our case we
take $\tau_\alpha = 0$ as the necessary and sufficient condition
of local geometric compatibility in each $B_\alpha$, 
which is equivalent to local holonomicity of the affine frame fields.
It is essential to observe that the
above condition is independent of the choice of a representative in the
equivalence class (\ref{eqvgGa}). In consequence, the local geometric
(in)compatibility characterizes the field (\ref{eqfBLa}).

     For a simply-connected $B$ the geometric compatibility implies
that the distortion field $f$ can be derived locally from a
displacement field $u$. If $u\rightarrow 0$ smoothly,
then $f\rightarrow {\rm const}$ along a path of geometrically
compatible distortion states.

     Consider now a multiply-connected $B$. In spite of vanishing
$\tau$, the affine connection may have non-trivial holonomy
for non-contractible loops, expressed by elements $h\ \epsilon\ \Gamma$.
For a given distortion field $f$ a loop from $B$ is mapped
into a loop in $\Lambda$ which, in turn, can be (uniquely) lifted
to a path in $G$ which starts from the
unit element $e\ \epsilon\ G$ and terminates in a certain (Burgers)
element $g\ \epsilon\ G$. It is straightforward to verify
that for geometrically compatible distortion fields $f$,
the Burgers element $g$ coincides with the holonomy element $h$.

     Apart from the teleparallelism connection, one can also
consider the Riemann--Cartan connection associated with a given
frame field. By straightforward calculation 
one can verify that the geometric compatibility condition can be
formulated in an alternative, although equivalent, way:
the associated teleparallelism
coincides with the associated Riemann--Cartan
connection.

   Note that the above stated criteria for geometric compatibity
are directly valid also for higher dimensionalities
and more general manifolds. In the next section we shall show
the existence of distortion fields in $E^2$ which are,
at the same time, geometrically compatible and topologically
non--trivial.

\section{Existence of compatible representatives}\indent

   Let us note that eqn. (\ref{eqsetB}) implies the existence of
a finite open covering
of B by contractible sets $B_\alpha$. For such a covering,
if the distortion field $f$ is compatible then each
$g_\alpha$ is derivable from an immersion
$w_\alpha: B_\alpha \rightarrow R^2$. Note, however, that
this immersion need not be an embedding.

    Now we can ask the question: do geometrically compatible
and toplogically non--trivial distortions exist? To show that
the answer is in the affirmative, consider the following
multi--valued analytic function

\equln{w(z) = {\frac{1}{2\pi i}}
           \sum_{\iota\,\epsilon\,J}
           [(a_\iota(z-z_\iota) + c_\iota)
           \log (z - z_\iota)
           - (b_\iota(z-z_\iota)^* + d_\iota)
           \log (z - z_\iota)^*]}{wfun}
where $z_\iota\ \epsilon\ A_\iota$.
For an appropriate covering
$\{B_\alpha\}$ this function translates itself into a collection of
immersions $w_\alpha: B_\alpha \rightarrow R^2$. This collection
defines a geometrically compatible distortion field provided
that $a_\iota$, $b_\iota$, and $c_\iota+d_\iota$ are complex
integers such that
\equn{\mid a_\iota\mid^2-\mid b_\iota\mid^2\ =\ 1.}
In consequence, the
holonomy element $h_\iota$ associated with each
$\iota\ \epsilon\ J$ equals the affine transformation
defined by
\equn{{\rm Re}\left( \begin{array}{rr}
       (a_\iota+b_\iota) & i(a_\iota-b_\iota)\\
       -i(a_\iota+b_\iota) & (a_\iota-b_\iota)
       \end{array}\right)
       \left(\begin{array}{l}
       x-x_\iota \\ y-y_\iota
       \end{array}\right)\ +\ \left( \begin{array}{r}
       {\rm Re}(c_\iota+d_\iota)\\ {\rm Im}(c_\iota+d_\iota)
       \end{array}\right),}
so that any $\gamma_\iota\ \epsilon\ \Gamma$
can be obtained in this way.
The group identity is realized only by
\equn{a_\iota\,=\,1,\ b_\iota\,=\,c_\iota+d_\iota\,=\,0}
so that all the remaining elements correspond to non--trivial
homotopy classes. Moreover, the immersions
$w_\alpha: B_\alpha \rightarrow R^2$ can be made embeddings
by a finite refinement of the covering $\{B_\alpha\}$.

   The above formulation allows us to draw also a contrasting
conclusion concerning the isometric approach to crystalline defects.
When in (\ref{LaGGa}) the group $G$ is restricted to the
Euclidean isometries of the plane, and $\Gamma$ is restricted
to the crystallographic
space group of the reference lattice, then the frames (\ref{framesr})
can be chosen orthogonal. In that case the associated metrics
can be represented by the unit matrix, what makes the Christoffel
symbols vanish for compatible configurations. In consequence
the teleparallelism connection coefficients also vanish,
and only homogeneous states survive. As a result,
in the isometric approach the geometrically
compatible distortions are always topologically trivial.
This is in contrast with
the situation in the affine theory, which admits a rich variety of
topological, non-trivial distortion fields.

    The above results can further be strenghtened by taking
\equn{w(z)\,=\,w_1(z,z^*) + w_2(z,z^*)}
where $w_1(z,z^*)$ is given by eqn. (\ref{wfun}), and 
the function $w_2(z, z^*)$ is, for every $z\,\epsilon\,B$,
holomorhic with respect to $z$
and anti--holomorphic with respect to $z^*$.
By appropriate choice of meromorphic terms in $w_2(z,z^*)$
one obtains geometrically compatible distortion fields
characterized by non--zero winding numbers.

\end{document}